\documentstyle[12pt]{article}
\newlength{\bredde}
\def\slash#1{\settowidth{\bredde}{$#1$}\ifmmode\,\raisebox{.15ex}{/}
\hspace*{-\bredde} #1\else$\,\raisebox{.15ex}{/}\hspace*{-\bredde} #1$\fi}
\newcommand{\beq}{\begin{equation}}
\newcommand{\eeq}{\end{equation}}
\newcommand{\beqn}{\begin{eqnarray}}
\newcommand{\eeqn}{\end{eqnarray}}

\def\gtwid{\raise.3ex\hbox{$>$\kern-.75em\lower1ex\hbox{$\sim$}}}
\def\ltwid{\raise.3ex\hbox{$<$\kern-.75em\lower1ex\hbox{$\sim$}}}
\newcommand{\noi}{\vspace{12pt}\noindent}
\newcommand{\lG}{\raise.3ex\hbox{$\stackrel{\leftarrow}{G}$}}
\newcommand{\lU}{\raise.3ex\hbox{$\stackrel{\leftarrow}{U}$}}
\newcommand{\lP}{\raise.3ex\hbox{$\stackrel{\leftarrow}{{\cal P}}$}}
\newcommand{\leta}{\raise.3ex\hbox{$\stackrel{\leftarrow}{\eta}$}}
\newcommand{\lOmega}{\raise.3ex\hbox{$\stackrel{\leftarrow}{\Omega}$}}
\newcommand{\ldr}{\raise.3ex\hbox{$\stackrel{\leftarrow}{\delta^r}$}}
\begin{document}
\title{Symmetries and the Antibracket: The Batalin-Vilkovisky Method}
\author{Jorge Alfaro\\
Facultad de F\'\i sica, Universidad Cat\'olica de Chile, \\
Casilla 306, Santiago 22\\
 CHILE}
\maketitle

\section{Introduction}

The most general quantization prescription available today is provided by the Batalin-Vilkovisky
(BV) method\cite{Batalin,review}. It includes all the results of the BRST method, and also permits a systematic
quantization of systems with an open algebra of constraints. Moreover it possesses a rich algebraic
structure, that have been considered in the description of String Field Theory \cite{strings,Verlinde}.

Because of the many advantages of the BV method, it is useful to understand more deeply the nature
and reasons for the different prescriptions that appear there. This we have done in a series of 
papers\cite{aldam1,aldam2,aldam3,aldam4}:
Impossing that the most general Schwinger -Dyson equations of a theory should come from a symmetry
principle(SD-BRST symmetry), we have been able to derive the BV method from standard BRST Lagrangian
quantization. In so doing, we have understood the meaning of the antifields, the appearence of
the graded canonical structure generated by the antibracket and ultimately, the underlying reason for
the existence of a master equation. From this pesrpective, it is quite natural to generalize the 
BV structure, first to non-abelian invariances of the path integral measure\cite{aldam2}, later to the most
general open algebra\cite{aldam3}\cite{aldam4}.

In these lectures, we review these recent developments: In section II, we study the derivation of BV from
Schwinger-Dyson-BRST symmetry, in section III, we consider some generalizations of the BV structure suggested
by our approach. In section IV, we present a connection between the Poisson bracket and the antibracket and
use it to relate the Nambu bracketc\cite{Nambu} with the generalized $n$-brackets of section III.
Finally in section V, we draw some conclusions.

\section{Derivation of the Batalin-Vilkovisky Method from Schwinger-Dyson BRST
Symmetry}

In this section, we review the derivation of the BV method from Schwinger-Dyson BRST symmetry
done in \cite{aldam1}.

Let us recapitulate the basic ingredients of the Batalin-Vilkovisky(BV) method
\cite{Batalin,review}.

Begin
with a set of fields $\phi^A(x)$ of given Grassmann parity
(statistics) $\epsilon(\phi^A)=\epsilon_A$, and then introduce for
each field a corresponding antifield $\phi^*_A$ of opposite Grassmann
parity $\epsilon(\phi^*_A)=\epsilon_A+1$.
The fields and antifields are taken to be canonically
conjugate,
\beq
(\phi^A,\phi^*_B) = \delta^A_B~,~~~ (\phi^A,\phi^B) =
(\phi^*_A,\phi^*_B) ~=~ 0 ,
\eeq
within a certain graded bracket structure $(\cdot,\cdot)$,
the antibracket:
\beq
(F,G) = \frac{\delta^r F}{\delta\phi^A}\frac{\delta^l G}
{\delta\phi^*_A}
-\frac{\delta^r F}{\delta\phi^*_A}\frac{\delta^l G}{\delta\phi^A}~.
\eeq
The subscripts $l$ and $r$ denote left and right differentiation,
respectively. The summation over indices $A$ includes an integration
over continuous variables such as space-time $x$, when required.

This antibracket is statistics-changing in the sense that
\beq
\epsilon[(F,G)] = \epsilon(F) + \epsilon(G) + 1 ,
\eeq
and satisfies the following exchange relation:
\beq
(F,G) ~=~ -(-1)^{(\epsilon(F)+1)(\epsilon(G)+1)}(G,F) ~.
\eeq
Furthermore, one may verify that the antibracket acts as a derivation
of the kind
\begin{eqnarray}
(F,GH) &=& (F,G)H + (-1)^{\epsilon(G)(\epsilon(F)+1)}G(F,H) \cr
(FG,H) &=& F(G,H) + (-1)^{\epsilon(G)(\epsilon(H)+1)}(F,H)G  ~,
\end{eqnarray}
and satisfies a Jacobi identity of the form
\beq
(-1)^{(\epsilon(F)+1)(\epsilon(H)+1)}(F,(G,H)) +
{\mbox{\rm cyclic perm.}} = 0 ~.
\eeq
Some simple consequences of these relations are that $(F,F) = 0$ for
any Grassmann odd $F$, and $(F,(F,F)) = ((F,F),F) = 0$ for any $F$.

The antifields $\phi^*_A$ are also given definite ghost numbers
$gh(\phi^*_A)$, related to those of the fields $\phi^A$:
\beq
gh(\phi^*_A) = - gh(\phi^A) - 1 ~.
\eeq
The absolute value of the ghost number can be fixed by requiring that
the action carries ghost number zero.

The Batalin-Vilkovisky quantization prescription can now be formulated
as follows. First solve the equation
\beq
\frac{1}{2}(W,W) ~=~ i\hbar \Delta W \label{QME}
\eeq
where
\beq
\Delta ~=~ (-1)^{\epsilon_A+1}\frac{\delta^r}{\delta\phi^A}
\frac{\delta^r}{\delta\phi^*_A} ~.\label{D}
\eeq
This $W$ will be the ``quantum action", presumed expandable in powers of
$\hbar$:
\beq
W = S + \sum_{n=0}^{\infty} \hbar^n M_n , \label{QA}
\eeq
and a boundary condition is that $S$ in eq. (\ref{QA}) should coincide with
the classical action when all antifields are removed, $i.e.$, after
setting $\phi^*_A = 0$. One can solve for the additional $M_n$-terms
through a recursive procedure, order-by-order in an $\hbar$-expansion.
To lowest order in $\hbar$ this is the Master Equation:
\beq
(S,S) ~=~ 0 ~.\label{ME}
\eeq
Similarly, one can view eq. (\ref{QME}) as the full ``quantum Master Equation".

A correct path integral prescription for the
quantization of the classical theory $S[\phi^A,\phi^*_A=0]$ is that
one should find an appropriate ``gauge fermion" $\Psi$ such that the partition
function is given by
\beq
{\cal{Z}} ~=~ \int [d\phi^A][d\phi^*_A]\delta(\phi^*_A -
\frac{\delta^r \Psi}{\delta \phi^A})\exp\left[\frac{i}{\hbar}W\right]~.\label{Z}
\eeq
This prescription guarantees gauge-independence of the $S$-matrix of
the theory. The ``extended action" $S \equiv S_{ext}[\phi^A,\phi^*_A]$
is a solution of the Master Equation (\ref{ME}),
and can been given by an expansion in powers of antifields
\cite{Batalin}. After the elimination of the antifields by the
$\delta$-function constraint in eq. (\ref{Z}), one can verify that the
action is invariant under the usual BRST symmetry, which we will
here denote by $\delta$.

In the following, we will uncover and derive in a simple manner the Batalin-Vilkovisky
formalism starting from two basic ingredients: A) Standard BRST Lagrangian
quantization, and B) The requirement that the most general Schwinger-Dyson
equations of the full quantum theory follow from a symmetry principle.
We shall throughout, unless otherwise stated, assume that when
ultraviolet regularization is required, a suitable regulator
which preserves the relevant BRST symmetry exists.

Schwinger-Dyson equations will thus play a crucial r\^{o}le in this
analysis. The idea is to enlarge the usual BRST
symmetry in precisely such a way that both the usual gauge-symmetry Ward
Identities {\em and} the most general Schwinger-Dyson equations both
follow from the same BRST Ward Identities. The way to do this is known
\cite{us};
it is a special case of collective field transformations that can
be used to gauge arbitrary symmetries \cite{us1}.

\subsection{No gauge Symmetries}

Consider a quantum field theory based on an action $S[\phi^A]$ without
any internal gauge symmetries. Such a quantum theory can be described
by a path integral.
\beq
{\cal{Z}} ~=~ \int [d\phi^A] \exp\left[\frac{i}{\hbar} S[\phi^A]\right] ,
\eeq
and the associated generating functional. 

Equivalently, such a quantum field theory can be entirely
described by the solution of the corresponding Schwinger-Dyson
equations, once appropriate boundary conditions have been imposed.
At the path integral level they follow
from invariances of the measure. Let us for simplicity consider the
case of a flat measure which is invariant under arbitrary local shifts,
$\phi^A(x) \to \phi^A(x) + \varepsilon^A(x)$. We can gauge this symmetry
by means of collective fields $\varphi^A(x)$: Suppose we transform the
original field as
\beq
\phi^A(x) \to \phi^A(x) - \varphi^A(x) ,
\eeq
then the transformed action $S[\phi^A-\varphi^A]$ is trivially invariant
under the local gauge symmetry
\beq
\delta \phi^A(x) = \Theta(x)~,~~~~ \delta \varphi^A(x) = \Theta(x) ,
\eeq
and the measure for $\phi^A$ in eq. (13) is also invariant. 
The gauge invariant functions are of the form $F[\phi^A-\varphi^A]$, for
any $F$.

We next
integrate over the collective field in the transformed path integral,
using the same flat measure.
The integration is of course very formal since it will include the whole
volume of the gauge group.\footnote{This situation is no different from
usual path integral manipulations of
gauge theories.} To cure this problem, we gauge-fix in the standard
BRST Lagrangian manner \cite{Kugo1}. That is, we add to the transformed
Lagrangian a BRST-exact term in such a way that the local gauge symmetry
is broken. In this case an obvious BRST multiplet consists of a
ghost-antighost pair $c^A(x), \phi^*_A(x)$, and a Nakanishi-Lautrup
field $B_A(x)$:
\begin{eqnarray}
\delta \phi^A(x) &=& c^A(x) \cr
\delta \varphi^A(x) &=& c^A(x) \cr
\delta c^A(x) &=& 0 \cr
\delta \phi^*_A(x) &=& B_A(x) \cr
\delta B_A(x) &=& 0 .
\end{eqnarray}

No assumptions will be made as to whether $\phi^A$ are of odd or even
Grassmann parity. We assign the usual ghost numbers to the new fields,
\beq
gh(c^A) = 1~,~~~ gh(\phi^*_A) = -1~,~~~ gh(B_A) = 0,
\eeq
and the operation $\delta$ is statistics-changing. The rules for
operating with $\delta$ are given in the Appendix.

Let us choose to gauge-fix the transformed action
by adding to the Lagrangian a term of the form
\beq
-\delta[\phi^*_A(x)\varphi^A(x)] = (-1)^{\epsilon(A)+1}B_A(x)
\varphi^A(x) - \phi^*_A(x)c^A(x) .
\eeq
The partition function is now again well-defined:
\beq
{\cal{Z}} ~=~ \int [d\phi][d\varphi][d\phi^*][dc][dB]
\exp\left[\frac{i}{\hbar}\left(S[\phi-\varphi] - \int dx\{
(-1)^{\epsilon(A)}B_A(x)\varphi^A(x) + \phi^*_A(x)c^A(x)\}
\right)\right] .
\eeq
Since the collective field has just been gauge fixed to zero, it
may appear useful to integrate both it and the field $B_A(x)$ out.
We are then left with
\begin{eqnarray}
{\cal{Z}} &=& \int [d\phi^A][d\phi^*_A][dc^A] \exp\left[\frac{i}{\hbar}
S_{ext}\right] \cr
S_{ext} &=& S[\phi^A] - \int dx \phi^*_A(x)c^A(x) ~,
\end{eqnarray}
which obviously coincides with the original expression (13) apart from
the trivially decoupled ghosts. But the remnant BRST symmetry is
still non-trivial: We find it in the usual way by substituting
for $B_A(x)$ its equation of motion. This gives
\begin{eqnarray}
\delta \phi^A(x) &~=~& c^A(x) \cr
\delta c^A(x) &~=~& 0 \cr
\delta \phi^*_A(x) &~=~& - \frac{\delta^l S}{\delta \phi^A(x)} .\label{SD}
\end{eqnarray}

The functional measure is also invariant under this symmetry,
according to our assumption about the measure for $\phi^A$, and
assuming a flat measure for $\phi^*_A$ as well.
The Ward Identities following from this symmetry are the
seeked-for Schwinger-Dyson equations:

$$
0 = \langle \delta\{\phi^*_A(x)F[\phi^A]\}\rangle
$$
, where
we have chosen $F$ to depend only on $\phi^A$ just to ensure that
the whole object carries overall ghost number zero. After integrating over both
ghosts $c^A$ and antighosts $\phi^*_A$, this Ward Identity can be
written
\beq
\langle \frac{\delta^lF}{\delta\phi^A(x)} +
\left(\frac{i}{\hbar}\right)\frac{\delta^lS}{\delta\phi^A(x)}
F[\phi^A] \rangle = 0 ~,
\eeq
that is, precisely the most general Schwinger-Dyson equations for
this theory. The symmetry (\ref{SD}) can be viewed as the BRST Schwinger-Dyson
algebra. 

Consider now the equation that expresses BRST invariance of the extended
action $S_{ext}$:
\begin{eqnarray}
0 = \delta S_{ext} &=& \int dx \frac{\delta^r S_{ext}}{\delta\phi^A(x)}
c^A(x) - \int dx \frac{\delta^r S_{ext}}{\delta\phi^*_A(x)}
\frac{\delta^l S}{\delta\phi^A(x)} \cr
&=& \int dx \frac{\delta^r S_{ext}}{\delta\phi^A(x)} c^A(x)
- \int dx \frac{\delta^r S_{ext}}{\delta\phi^*_A(x)}
\frac{\delta^l S_{ext}}{\delta\phi^A(x)}~.
\end{eqnarray}
In the last line we have used the fact that $S$ differs from
$S_{ext}$ by a term independent of $\phi^A$. Using the notation of
the antibracket (2), this is seen to correspond to a Master Equation
of the form
\beq
\frac{1}{2}(S_{ext},S_{ext}) = - \int dx \frac{\delta^r S_{ext}}
{\delta\phi^A(x)}c^A(x) ~.
\eeq

The ghosts $c^A$ play the r\^{o}le of spectator fields in the
antibracket. But their appearance on the r.h.s. of the Master Equation
ensures that the solution $S_{ext}$ will contain these fields.

The extended action of Batalin and Vilkovisky does however
not coincide with $S_{ext}$ as defined above. 
But suppose we integrate {\em only} over these ghosts $c^A(x)$,
without integrating over the corresponding antighosts $\phi^*_A(x)$.
Then the partition function reads
\beq
{\cal{Z}} = \int [d\phi^A][d\phi^*_A]\delta\left(\phi^*_A\right)
\exp\left[\frac{i}{\hbar}S[\phi^A]\right] .
\eeq

What has happened to the BRST algebra?
For the present case of a ghost field $c^A$
appearing linearly in the action before being integrated out, we
can derive the correct substitution rule as follows.
First, we should really phrase the question in a more precise
manner. What we need to know is how to replace $c$ inside
the path integral, $i.e.$, inside Green functions.
This will automatically give us the correct transformation
rules for those fields that are not integrated out. Consider
the identity
\beq
\int [dc] F(c^B(y))\exp\left[-\frac{i}{\hbar}\int dx
\phi^*_A(x)c^A(x)
\right] = F\left(i\hbar\frac{\delta^l}{\delta \phi^*_B(y)}\right)
\exp\left[-\frac{i}{\hbar}\int dx\phi^*_A(x)c^A(x)\right] ~.\label{28}
\eeq

Eq. (\ref{28}) teaches us that it is not enough to replace $c$ by its
equation of motion ($c(x)=0$); a ``quantum correction" in the
form of the operator $\hbar\delta/\delta\phi^*$ must be added as
well. The appearance of this operator is the final step
towards unravelling the canonical structure in the formalism of
Batalin and Vilkovisky. It also shows that even in this trivial
case we have to include ``quantum corrections" to BRST symmetries
if we insist on integrating out only one ghost field, while
keeping its antighost.

It is important that the operator $\hbar\delta^l/\delta\phi^*$
in eq. (\ref{28}) always acts on the integral (really a $\delta$-function)
to its right.

We now make this replacement, having always in mind that it is only
meaningful inside the path integral. For the BRST transformation
itself we get, upon one partial integration
\begin{eqnarray}
\delta\phi^A(x) &=& i\hbar(-1)^{\epsilon_A}
\frac{\delta^r}{\delta\phi^*_A(x)} \cr
\delta\phi^*_A(x) &=& - \frac{\delta^l S}{\delta\phi^A(x)} ~.
\end{eqnarray}
We know from our derivation that this transformation leaves at least
the combination of measure and action invariant. As a check, if we
consider the same Ward Identity as above, based on $0 = \langle
\delta\{\phi^*_A(x)F[\phi^A]\}\rangle$, we recover the
Schwinger-Dyson equation (22).

The original $S[\phi^A]$ can in this case be identified with the
extended action of Batalin and Vilkovisky, and the antighost
$\phi^*_A$ is the antifield corresponding to $\phi^A$. Because
there are no internal gauge symmetries, the extended action turns
out to be independent of the antifields. Although our $S_{ext}$
of eq. (20) cannot
be identified with the extended action of Batalin-Vilkovisky, the
Master Equation derived in eq. (23) is of course very similar
to their corresponding Master Equation.
Writing eq.(23) in terms of the antibracket is a little forced. 
It is done in order to facilitate the
comparison.

Finally, it only remains to be seen what has happened to this
bracket structure after having integrated out the ghost. We will
keep the same notation as before, so that in
this case the ``extended" action is trivially equal to the
original action: $S_{ext} = S[\phi^A]$. Let us then again
consider the variation of an arbitrary functional $G$, this time
only a function of $\phi^A$ and $\phi^*_A$. Inside the path
integral (and only there!) we can represent the variation of $G$
as:
\beq
\delta G[\phi^A,\phi^*_A] = \int dx\frac{\delta^r G}{\delta\phi^A(x)}
\left[\frac{\delta^l S_{ext}}{\delta\phi^*_A(x)} +
(i\hbar)(-1)^{\epsilon_A}\frac{\delta^r}
{\delta\phi^*_A(x)}\right] - \int dx\frac{\delta^r G}
{\delta\phi^*_A(x)}\frac{\delta^l S_{ext}}{\delta\phi^A(x)}~,
\eeq
where the derivative operator no longer acts on the
$\delta$-function of $\phi^*_A$. We have kept the term proportional
to $\delta^l S_{ext}/\delta\phi^*_A$, even though $S_{ext}$ in this
simple case is independent of $\phi^*_A$ (and that term therefore
vanishes). It comes from the partial integration with respect to
the operator $\delta^l/\delta^*_A$, and is there in general when
$S_{ext}$ depends on $\phi^*_A$.

This equation precisely describes the ``quantum deformation" of the
classical BRST charge, as it occurs in the Batalin-Vilkovisky
framework:
\beq
\delta G = (G,S_{ext}) - i\hbar \Delta G ~.
\eeq
The BRST operator in this form is often denoted by $\sigma$.

We have here introduced
\beq
\Delta \equiv (-1)^{\epsilon_A+1}\frac{\delta^r\delta^r}
{\delta\phi^*_A(x)
\delta\phi^A(x)}
\eeq
which is identical to the operator (\ref{D}) of Batalin and
Vilkovisky. Again, this term arises as a consequence of the
partial integration which allows us to expose the operator
$\delta/\delta\phi^*_A(x)$ that otherwise acts only on the
$\delta$-functional $\delta(\phi^*_A)$. In other words, the
$\delta$-function constraint on $\phi^*_A(x)$ is now considered as part
of the functional measure for $\phi^*_A$.

Precisely which properties of the partially-integrated extended action
are then responsible for the canonical structure behind the
Batalin-Vilkovisky formalism? As we have seen, the crucial ingredients
come from integrating out the Nakanishi-Lautrup fields $B_A$ and the
ghosts $c^A$. Integrating out $B_A$ changes the BRST variation of the
antighosts $\phi^*_A$ into $-\delta^l S_{ext}/\delta\phi^A$.
{\em This is
an inevitable consequence of introducing collective fields as shifts
of the original fields (and hence enforcing Schwinger-Dyson equations),
and then gauge-fixing them to zero}. But this only provides half of the
canonical structure, making, loosely speaking, $\phi^A$ canonically
conjugate to $\phi^*_A$, but not vice versa. The rest is provided by
integrating over the ghosts $c^A$; at the linear level it changes the
BRST variation of the fields $\phi^A$ themselves
into $\delta^l S_{ext}/\delta\phi^*_A$. {\em This in turn is again an
inevitable consequence of having introduced the collective fields as
shifts, and then having gauge-fixed them to zero.}\footnote{Gauge-fixing
the collective fields to zero implies linear couplings to
the auxiliary fields and ghosts, respectively. This is one of the central
properties of the extended action that leads to the canonical structure,
and to the fact that the extended action $S_{ext}$ itself is the
(classical) BRST generator. However, this does not exclude the
possibility that different gauge fixings of the shift symmetries could
produce a more general formalism.}

The latter operation is,
however, complicated by the fact that the fields $\phi^*_A$ which are
fixed in the process of integrating out $c^A$ are chosen to remain
in the path integral. This makes it impossible to discard the ``quantum
correction" to $\delta S_{ext}/\delta\phi^*_A$. So, in fact, if one
insists on keeping these antighosts $\phi^*_A$, now seen as canonically
conjugate partners of $\phi^A$, the simple canonical structure is in this
sense never truly realized.

We have seen these features only in what is the trivial
case of no internal gauge symmetries. But as we shall show in the
following  section, they hold in greater generality.

\subsection{Gauge Theories: Yang-Mills}

Let us illustrate our approach with another simple example: Yang-Mills theory. 

We start with the pure Yang-Mills action $S[A_\mu]$, and define a
covariant derivative with respect to $A_\mu$ as
\beq
D^{(A)}_\mu ~\equiv~ \partial_\mu - [A_\mu,~] .
\eeq
Next, we introduce the collective field $a_\mu$ which will enforce
Schwinger-Dyson equations as a result of the BRST algebra:
\beq
A_\mu(x) \to A_\mu(x) - a_\mu(x) .
\eeq
In comparison with the previous example, the only new aspect
here is that the transformed action $S[A_\mu-a_\mu]$ actually has two
independent gauge symmetries. Because of the redundancy introduced by
the collective field, we can write the two symmetries in different ways.
To make contact with the Batalin-Vilkovisky formalism, we will choose
the following version, 
\begin{eqnarray}
\delta A_\mu(x) &=& \Theta_\mu(x) \cr
\delta a_\mu(x) &=& \Theta_\mu(x) - D^{(A-a)}_\mu\varepsilon(x) ,
\end{eqnarray}
which also shows the need for being careful in defining what we mean
by a covariant derivative. (The general principle is that we choose
the original gauge symmetry of the original field to be carried
entirely by the collective field; the transformation of the original
gauge field is then always just a shift.)
Although $\Theta(x)$ includes arbitrary
deformations, it only leaves the transformed $\em field$ invariant,
while of course the action is also invariant under Yang-Mills gauge
transformations of this transformed field itself. Hence the need for
including two independent gauge transformations.

We now gauge-fix these two gauge symmetries, one at a time, in the
standard BRST manner. As a start, we introduce a suitable
multiplet of ghosts and auxiliary fields. We need one Lorentz
vector ghost $\psi_\mu(x)$ for the shift symmetry of $A_\mu$,
and one Yang-Mills ghost $c(x)$. These are of course Grassmann
odd, and both carry the same ghost number
\beq
gh(\psi_\mu) ~=~ gh(c) ~=~ 1 .
\eeq

Next, we gauge-fix the shift symmetry of $A_\mu$ by removing the
collective field $a_\mu$. This leads us to introduce a corresponding
antighost $A^*_\mu(x)$, Grassmann odd, and an auxiliary field
$b_\mu(x)$, Grassmann even. They have the usual ghost number
assignments,
\beq
gh(A^*_\mu) ~=~ -1~,~~~~ gh(b_\mu) ~=~ 0 ~,
\eeq
and we now have the nilpotent BRST algebra
\begin{eqnarray}
\delta A_\mu(x) &=& \psi_\mu(x) \cr
\delta a_\mu(x) &=& \psi_\mu(x) - D^{(A-a)}_\mu c(x) \cr
\delta c(x) &=& -\frac{1}{2}[c(x),c(x)] \cr
\delta \psi_\mu(x) &=& 0 \cr
\delta A^*_\mu(x) &=& b_\mu(x) \cr
\delta b_\mu(x) &=& 0 .
\end{eqnarray}

Fixing $a_\mu(x)$ to zero is achieved by adding a term
\beq
-\delta[A^*_\mu(x)a^\mu(x)] = - b_\mu(x)a^\mu(x) -
A^*_\mu(x)\{\psi^\mu - D_{(A-a)}^\mu c(x)\}
\eeq
to the Lagrangian. We shall follow the usual rule of only starting
to integrate over
{\em pairs} of ghost-antighosts in the partition function. With this
rule we shall still keep $c(x)$ unintegrated (since we have not yet
introduced its corresponding antighost), but we can now integrate
over both $\psi_\mu(x)$ and $A^*_\mu(x)$. This leads to the following
extended, but not yet fully gauge-fixed, action $S_{ext}$:
\begin{eqnarray}
{\cal{Z}} &=& \int [dA_\mu][da_\mu][d\psi_\mu][dA^*_\mu][db_\mu]
\exp\left[\frac{i}{\hbar}S_{ext}\right] \cr
S_{ext} &=& S[A_\mu-a_\mu] - \int dx\{b_\mu(x)a_\mu(x) +
A^*_\mu(x)[\psi^\mu(x) - D_{(A-a)}^\mu c(x)]\}
\end{eqnarray}
This extended action is invariant under the BRST transformation (36).
The full integration measure is also invariant. Of course, the expression
above is still formal, since we have not yet gauge fixed ordinary
Yang-Mills invariance. 

Furthermore, we are eventually going to integrate over the ghost
$c$, which already now appears in the extended action. If we insist
that the Schwinger-Dyson equations involving this field, $i.e$,
equations of the form
\beq
0 = \int [dc] \frac{\delta^l}{\delta c(x)}\left[F e^{\frac{i}{\hbar}
\left[\mbox{\rm Action}\right]}\right]
\eeq
are to be satisfied automatically (for reasonable choices of functionals
$F$) by means of the full unbroken BRST
algebra, we must introduce yet one more collective field. This
new collective field, call it $\tilde{c}(x)$, is Grassmann odd, and
has $gh(\tilde{c}) = 1$. Now shift the Yang-Mills ghost:
\beq
c(x) \to c(x) - \tilde{c}(x) .
\eeq
To fix the associated fermionic gauge symmetry, we introduce a new
BRST multiplet of a ghost-antighost pair and an auxiliary field. We
follow the same rule as before and let the transformation of the new
collective field $\tilde{c}$ carry the (BRST) transformation of the
original ghost $c$, $viz.$,
\begin{eqnarray}
\delta c(x) &=& C(x) \cr
\delta \tilde{c}(x) &=& C(x) + \frac{1}{2}[c(x)-\tilde{c}(x),
c(x)-\tilde{c}(x)] \cr
\delta C(x) &=& 0 \cr
\delta c^*(x) &=& B(x) \cr
\delta B(x) &=& 0 .
\end{eqnarray}

It follows from (41) that ghost number assignments should be as
follows:
\beq
gh(C) = 2~,~~~ gh(c^*) = -2~,~~~ gh(B) = -1 ,
\eeq
and $B$ is a fermionic Nakanishi-Lautrup field. 
Next, gauge-fix $\tilde{c}(x)$ to zero by adding a term
\beq
-\delta[c^*(x)\tilde{c}(x)] =  B(x)\tilde{c}(x) -
c^*(x)\{C(x) + \frac{1}{2}[c(x)-\tilde{c}(x),c(x)-\tilde{c}(x)]\}
\eeq
to the Lagrangian. This gives us the fully extended action,
\begin{eqnarray}
S_{ext} &=& S[A_\mu-a_\mu] - \int dx\{b_\mu(x)a_\mu(x) +
A^*_\mu(x)[\psi^\mu(x) - D^\mu_{(A-a)}\{c(x)- \tilde{c}(x)\}] \cr
&& - B(x)\tilde{c}(x) + c^*(x)(C(x)+\frac{1}{2}
[c(x)-\tilde{c}(x),c(x)-\tilde{c}(x)])\} ,
\end{eqnarray}
with the partition function so far being integrated over all fields
appearing above, except for $c$ (whose antighost $\bar{c}$ still has
to be introduced when we gauge-fix the original Yang-Mills symmetry).

The extended action {\em and} the functional measure is formally
invariant under the following set of transformations:

\vspace{0.6cm}
$
\begin{array}{llllll}
\delta A_\mu(x) &=& \psi_\mu(x) , & \delta \psi_\mu(x)
&=& 0 , \\
\delta a_\mu(x) &=& \psi_\mu(x)-D^{(A-a)}_\mu[c(x)-\tilde{c}(x)]
, & \delta c(x)
&=&  C(x) ,\\
\delta A^*_\mu(x) &=& b_\mu(x) & \delta b_\mu(x) , &=& 0 ,\\
\delta \tilde{c}(x) , &=& C(x) +
\frac{1}{2}[c(x)-\tilde{c}(x),c(x)-\tilde{c}(x)]  &
\delta C(x) &=& 0, \\
\delta c^*(x) &=& B(x) & \delta B(x) &=& 0 ~.
\end{array}
$
\vspace{0.6cm}

As the notation indicates, the fields $A^*_\mu(x)$ and $c^*(x)$ can be
identified with the Batalin-Vilkovisky antifields of $A_\mu(x)$ and
$c(x)$, respectively. {\em These antifields are the usual antighosts
of the collective fields enforcing Schwinger-Dyson equations through
shift symmetries.}

Note that the general rule of assigning ghost number and Grassmann
parity to the antifields,
\beq
gh(\phi^*_A) = - (gh(\phi^A) + 1)~,~~~~~ \epsilon(\phi^*_A) =
\epsilon(\phi^A) + 1~,
\eeq
arises in a completely straightforward manner. Here, it is a
simple consequence of the fact that the BRST operator raises ghost
number by one unit (and changes statistics), supplemented with the
usual rule that antighosts have opposite ghost number of the ghosts.

The extended action (44) above has more fields than the extended
action of Batalin and Vilkovisky, and the transformation laws of what
we identify as antifields do not match those of ref. \cite{Batalin}.
In the form we have presented it, the BRST symmetry is nilpotent also
off-shell. If we integrate over the auxiliary fields $b_\mu$ and
$B$ (and subsequently over $a_\mu$ and $\tilde{c}$) the BRST symmetry
becomes nilpotent only on-shell. The extended action then takes the
following form:\footnote{We are dropping the subscript
on the covariant derivative since no confusion can arise at this
point.}
\beq
S_{ext} = S[A_\mu] - \int dx\{A^*_\mu(x)[\psi^\mu(x) -
D^\mu c(x)] + c^*(x)(C(x)+\frac{1}{2}[c(x),c(x)])\} .
\eeq
This differs from the extended action of Batalin and Vilkovisky by
the terms involving the ghost fields $\psi_\mu(x)$ and $c^*(x)$.
Comparing with the
case of no gauge symmetries, this is exactly what we should expect.
These ghost fields $\psi_\mu$ and $c^*$ ensure the correct
Schwinger-Dyson equations for $A_\mu$ and $c$, respectively.

To find the corresponding BRST symmetry we use
the equations of motion for the auxiliary fields $b_\mu$ and $B$, and
use the $\delta$-function constraints on $a_\mu$ and $\tilde{c}$. This
gives
\begin{eqnarray}
\delta A_\mu(x) &=& \psi_\mu(x) \cr
\delta\psi_\mu(x) &=& 0 \cr
\delta c(x) &=& C(x) \cr
\delta C(x) &=& 0 \cr
\delta A^*_\mu(x) &=& - \frac{\delta^l S_{ext}}{\delta A_\mu(x)} \cr
\delta c^*(x) &=& - \frac{\delta^l S_{ext}}{\delta c(x)}
\end{eqnarray}

BRST invariance of the extended action (46) immediately implies
that it satisfies a Master Equation we can write as
\beq
\int dx\frac{\delta^r S_{ext}}{\delta A^*_{\mu}(x)}
\frac{\delta^l S_{ext}}{\delta A^{\mu}(x)} + \int dx
\frac{\delta^r S_{ext}}{\delta c^*(x)}
\frac{\delta^l S_{ext}}{\delta c(x)}
= \int dx\frac{\delta^r S_{ext}}{\delta A_{\mu}(x)}\psi_{\mu}(x) +
\int dx\frac{\delta^r S_{ext}}{\delta c(x)}C(x) ~.
\eeq

Note that this Master Equation precisely is of the form
\beq
\frac{1}{2} (S_{ext},S_{ext}) = - \int dx \frac{\delta S_{ext}}
{\delta \phi^A(x)}c^A(x) ~,
\eeq
with two ghost fields $c^A$ that are just $\psi_{\mu}$ and
$C(x)$.

To finally make contact with the Batalin-Vilkovisky formalism, let us
integrate out the ghost $\psi_\mu$. As in eq.(\ref{28}), we shall use an
identity of the form
\begin{eqnarray}
&&\int[d\psi_\mu]F[\psi^\mu(y)]\exp\left[-\frac{i}{\hbar}\int dx
A^*_\mu(x)\{\psi^\mu(x) - D^\mu c(x)\}\right] \cr
&=&F\left[D^\mu c(y)+ (i\hbar)\frac{\delta^l}{\delta A^*_\mu(y)}\right]
\int [d\psi_\mu] \exp\left[-\frac{i}{\hbar}\int dx
A^*_\mu(x)\{\psi^\mu(x) - D^\mu c(x)\}\right]  \cr
&=&\exp\left[\frac{i}{\hbar}\int dx A^*_{\mu}(x)D^{\mu}c(x)\right]
F\left[(i\hbar)\frac{\delta^l}{\delta A^*_{\mu}(y)}\right]
\delta(A^*_{\mu})
\end{eqnarray}

This shows that we should replace $\psi_\mu(x)$ by its equation of
motion, plus the shown quantum correction of $\cal{O}(\hbar)$
which then acts on the rest of the integral, or equivalently, by
just the derivative operator (which then acts solely on the
functional $\delta$-function). To get a useful representation of
$\psi^{\mu}(y)$ we integrate the latter version of the identity
by parts, thus letting the derivative
operator act on everything except $\delta(A^*_{\mu})$. This
automatically brings down the equations of motion for $\psi^{\mu}$.

Having in this manner integrated out $\psi^{\mu}$ and $C$, the
partition function reads
\begin{eqnarray}
{\cal{Z}} &=& \int [dA_\mu][dA^*_\mu][dc^*]\delta(A^*_\mu)
\delta(c^*)\exp\left[\frac{i}{\hbar}S_{ext}\right] \cr
S_{ext} &=& S[A_\mu] + \int dx\{A^*_\mu(x)D^\mu c(x) -
\frac{1}{2}c^*[c(x),c(x)]\} ,
\end{eqnarray}
and the {\em classical} BRST symmetry follows as discussed above
by substituting only the equations of motion for $\psi_\mu$ and
$C$:
\begin{eqnarray}
\delta A_\mu(x) &=& \frac{\delta^l S_{ext}}{\delta A^*_{\mu}(x)}
= D_\mu c(x) \cr
\delta c(x) &=& \frac{\delta^l S_{ext}}{\delta c^*(x)}
-\frac{1}{2}[c(x),c(x)] \cr
\delta A^*_\mu(x) &=& -\frac{\delta^l S_{ext}}{\delta A_\mu(x)}
\cr \delta c^*(x) &=& -\frac{\delta^l S_{ext}}{\delta c(x)} ~.
\end{eqnarray}

This is the usual extended action of Batalin and Vilkovisky and the
corresponding classical BRST symmetry. Of course, in the partition
function the integrals over $A^*_\mu$ and $c^*$ are trivial. This
is as it should be, because by integrating out these antighosts we
should finally recover the starting measure and the still not
gauge-fixed Yang-Mills action. What we have provided here is thus
only a very precise functional derivation of the extended action. It
shows that we {\em can} understand the extended action in the usual
path integral framework, and that the integration measures for
$A^*_\mu$ and $c^*$ (with the accompagnying $\delta$-function
constraints) are provided automatically.

Obviously, the extended action is not yet very useful from the point
of view of ordinary BRST gauge fixing. To unravel some of the
mechanisms behind the Batalin-Vilkovisky scheme, it is nevertheless
advantageous to keep the antifields.
In fact, it is even more useful to return to the formulation in eq.
(46), where there is yet no split into a classical and a quantum part
of the symmetry. Let us therefore take (46) as the starting action,
and now just gauge-fix in a standard manner the Yang-Mills symmetry.
Choosing, $e.g.$, a covariant gauge, we therefore finally extend the
BRST multiplet to include a Yang-Mills antighost $\bar{c}$ and a
Nakanishi-Lautrup scalar $b$. For there to be no doubt, let us also
note that these fields have
\beq
gh(\bar{c}) = -1~,~~~~ gh(b) = 0 ~.
\eeq
The BRST transformations are the usual $\delta \bar{c}(x) = b(x)$
and $\delta b(x) = 0$.
Gauge fixing to a covariant $\alpha$-gauge can be achieved by adding
a term
\beq
\delta[\bar{c}(x)\{\partial_\mu A^\mu(x) - \frac{1}{2\alpha}b(x)\}] =
b(x)\partial_\mu A^\mu(x) + \bar{c}(x)\partial_\mu\psi^\mu(x)
- \frac{1}{2\alpha}b(x)^2
\eeq
to the Lagrangian. The corresponding completely gauge-fixed extended
action then reads
\begin{eqnarray}
S_{ext} &=& S[A_\mu] - \int dx\{A^*_\mu(x)[\psi^\mu(x) -
D^\mu c(x)] + c^*(x)(C(x)+\frac{1}{2}[c(x),c(x)]) \cr
&& - b(x)\partial_\mu
A^\mu(x) - \bar{c}(x)\partial_\mu\psi^\mu(x)
+ \frac{1}{2\alpha}b(x)^2\} .
\end{eqnarray}

Now integrate out $\psi^\mu$ and $C$. The result is indeed a
partition function of the form introduced by the formalism of
Batalin and Vilkovisky:
\begin{eqnarray}
{\cal{Z}} &=& \int [dA_\mu][dA^*_\mu][d\bar{c}][dc][dc^*][db]
\delta(A^*_\mu + \partial_\mu\bar{c})\delta(c^*)e^{\left[
\frac{i}{\hbar}S_{ext}\right]} \cr
S_{ext} &=& S[A_\mu] + \int dx\{A^*_\mu(x)D^\mu c(x)
- \frac{1}{2}c^*(x)[c(x),c(x)] \cr
&& +b(x)\partial_\mu A^\mu(x)
- \frac{1}{2\alpha}b(x)^2\} .
\end{eqnarray}

Note that by adding the Yang-Mills gauge-fixing terms, the
$\delta$-function constraint on the antifield $A^*_\mu$ has been shifted.
Thus when doing the $A^*_\mu$-integral, we are in effect substituting
not $A^*_\mu(x) = 0$ but
\beq
A^*_\mu(x) = -\partial_{\mu}\bar{c}(x) =
\frac{\delta^r \Psi}{\delta A^\mu(x)} ~,
\eeq
where $\Psi$ is defined as the term whose BRST variation is added to the
action, $i.e.$, in this particular case,
\beq
\Psi = \int dx\{\bar{c}(x)(\partial^\mu A_\mu(x) - \frac{1}{2\alpha}
b(x))\} ~.
\eeq
Upon doing the $A^*_\mu$ and $c^*$ integrals, one recovers the
standard covariantly gauge-fixed Yang-Mills theory
\beq
S = S[A_\mu] + \int dx\{\bar{c}(x)\partial^\mu D_\mu c(x) +
b(x)\partial^\mu A_\mu(x) - \frac{1}{2\alpha}b(x)^2\}.
\eeq

The identification (57) strongly suggests that one can see gauge
fixing as a particular canonical transformation involving new fields
(the antighosts $\bar{c}$). But there are terms in eq. (59) (those
involving $b(x)$) which
do not immediately follow from this perspective. In the
Batalin-Vilkovisky
framework, this is resolved by noting that one can always add
terms of new fields and antifields with trivial antibrackets.
In this version of the gauge-fixing procedure, one returns
to the (``minimally") extended action of eq. (56)
and extends it in a ``non-minimal" way.  In
the Yang-Mills case, this includes an additional term in the
action of the form
\beq
S_{nm} = \int dx \bar{c}^*(x)b(x) ~,
\eeq
with $\bar{c}^*$ and $b$ having the same ghost number and Grassmann
parity:
\beq
gh(\bar{c}^*) = gh(b) = 0 ~;~~~~~~\epsilon(\bar{c}^*) =
\epsilon(b) = 0 ~.
\eeq
As the notation indicates, these new fields $\bar{c}^*$ and $b$ are
indeed just the antifield of $\bar{c}$, and the usual
Nakanishi-Lautrup field, respectively.

Gauge fixing to the same gauge as above can then be achieved by
the same gauge fermion (58) which now affects both $A^*_{\mu}$ and
$\bar{c}^*$ within the antibracket.
It can thus be seen as the canonical transformation
that shifts $A^*_{\mu}$ and $\bar{c}^*$ from zero to
\beq
A^*_{\mu}(x) = \frac{\delta^r \Psi}{\delta A^{\mu}(x)} ~,~~~~~~
\bar{c}^*(x) = \frac{\delta^r \Psi}{\delta \bar{c}(x)} ~.
\eeq
Since $\Psi$ does not depend on the antifields, this canonical
transformation leaves all {\em fields} $A_{\mu}, c$ and $\bar{c}$
unchanged.

Can we understand the non-minimally extended action from our point
of view too? Consider the stage at which we introduce the antighost
$\bar{c}$. This field does not yet appear in the action, but we can
of course still introduce a corresponding collective ``shift" field
$\bar{c}'$ for $\bar{c}$
as well.
The corresponding BRST multiplet consists of a new ``shift-antighost"
$\lambda(x)$, an ``anti-antighost" $\bar{c}^*(x)$, and the associated
auxiliary field $B'(x)$:
\begin{eqnarray}
\delta \bar{c}(x) &=& \lambda(x) \cr
\delta \bar{c}'(x) &=& \lambda(x) - b(x) \cr
\delta \lambda(x) &=& 0 \cr
\delta b(x) &=& 0 \cr
\delta \bar{c}^*(x) &=& B'(x) \cr
\delta B'(x) &=& 0 ~.
\end{eqnarray}
The assignments will then have to be exactly as in eq. (61),
supplemented with $gh(B') = \epsilon(B') = 1$. We are again dealing
with {\em two} symmetries, because the shifted field $\bar{c}(x) -
\bar{c}'(x)$ itself can still be shifted by the usual Nakanishi-Lautrup
field. Let us now gauge-fix this large symmetry. We do
it in the most simple manner by adding a term
\beq
-\delta[\bar{c}^*(x)\bar{c}'(x)] = B'(x)\bar{c}'(x) - \bar{c}^*(x)(
\lambda(x) - b(x))
\eeq
to the Lagrangian. The integrals over $B'$ and $\bar{c}'$ are of course
trivial and we are left with the non-minimally extended
action for this theory plus, as expected, the corresponding term with
the new ghost $\lambda$. The final gauge-fixing of the Yang-Mills
symmetry will now consist in adding, instead of eq. (54),
\beq
\delta[\bar{c}(x)\{\partial_\mu A^\mu(x) - \frac{1}{2\alpha}
b(x)\}] =
\lambda(x)\partial_\mu A^\mu(x) + \bar{c}(x)\partial_\mu\psi^\mu(x)
- \frac{1}{2\alpha}\lambda(x)b(x) ~.
\eeq

Before Yang-Mills gauge fixing, the integral over $\lambda(x)$ just gave
a factor of $\delta(\bar{c}^*)$. After adding the gauge-fixing term,
this is modified:
\beq
\delta(\bar{c}^*) \to \delta(\bar{c}^*(x) - \partial^{\mu}A_{\mu}(x) +
\frac{1}{2\alpha}b(x))~.
\eeq
Substituting this back into the extended action, we recover the result (59).
Note that this indeed can be viewed as a canonical transformation within
the antibracket. All the correct $\delta$-function constraints are
provided by the collective fields and their ghosts. Since the functional
$\Psi$ has been chosen to depend only on the fundamental fields, and
not on the antifields, the fields $A_{\mu}, c$ and $\bar{c}$ are all
left untouched by this canonical transformation. Extending the action
from the minimal to the non-minimal case is equivalent to demanding that
also Schwinger-Dyson equations for $\bar{c}(x)$ follow as Ward identities
of the BRST symmetry. Since the antighost $\bar{c}$ remains in the path
integral after gauge fixing, it would indeed be very unnatural not to
demand that correct Schwinger-Dyson equations for this field follow as
well. As shown, this requirement automatically leads to the
{\em non-minimally} extended action.

As for the functional measure, we have stressed earlier that
we always assume the existence of a suitable regulator that
preserves the pertinent BRST symmetry. We can make this statement
a little more explicit by detailing the required symmetries of
the measure in this Yang-Mills case. Before integrating out any
fields, the measures for $A_{\mu}, c$ and $\bar{c}$ should all
be invariant under local shifts. For $A_{\mu}$ this corresponds
to the usual euclidean measure (and it is very difficult to imagine
this shift symmetry being broken by any reasonable regulator),
while for $c$ and $\bar{c}$ this is consistent with the usual rules
of Berezin integration. The measures for the three collective
fields should in addition be invariant under what corresponds to
usual Yang-Mills BRST transformations, a property that indeed holds
formally. Finally, the measures for all antifields are only
required to be invariant under local shifts. After having
integrated out the auxiliary fields $B_A$, invariance of these
measures of the antifields is now non-trivial but one can check
explicitly that it is formally satisfied. This should indeed
be the case, because it is straightforward to check that the
action remains invariant. Since at least the {\em combination}
of measure and action must remain invariant after integrating
out some of the fields, invariance of the measure is in this
case formally guaranteed.

Let us finally point out that once the antighost $\bar{c}$ is being
treated on equal footing with $A_{\mu}$ and $c$, a Master Equation
of the form
\beq
\frac{1}{2}(S_{ext},S_{ext}) = - \int dx\frac{\delta^r S_{ext}}
{\delta\phi^A(x)}c^A(x)
\eeq
now holds with $\phi^A$ denoting {\em all} the fields that finally
remain in the path integral: $A_{\mu},c$ and $\bar{c}$. Similarly,
the BRST algebra becomes, upon integrating out the collective fields
$\varphi^A$ and the auxiliary fields $B_A$, of the very simple form
(21) we encountered already in the case of no internal gauge
symmetries.

The general case of arbitrary gauge symmetries is considered in
\cite{aldam1}.

\subsection{Quantum Master Equations}

We have seen from the collective field method that for closed
irreducible gauge algebras we get an extended action that can
be split into a part independent of the new ghosts $c^A$, and
a simple quadratic term of the form $\phi^*_Ac^A$. Let us, for
reasons that will become evident shortly, denote the part which
is independent of $c^A$ by $S^{(BV)}, i.e.$:
\beq
S_{ext}[\phi,\phi^*,c] = S^{(BV)}[\phi,\phi^*] - \phi^*_Ac^A~.
\eeq

This action is invariant under the transformations
\begin{eqnarray}
\delta \phi^A &=& c^A \cr
\delta c^A &=& 0 \cr
\delta \phi^*_A &=& - \frac{\delta^l S_{ext}}
{\delta \phi^A} ~.
\end{eqnarray}
Moreover, the functional measure is also formally guaranteed to be
invariant in this case. It follows that in this case the Ward
Identities of the kind
$0 = \langle \delta[\phi^*_A F[\phi] \rangle$ are the most general
Schwinger-Dyson equations for the quantum theory defined by the
classical action $S[\phi]$.

But demanding that {\em both} the action $S_{ext}$ {\em and} the
functional measure be invariant under the BRST Schwinger-Dyson
symmetry above is not the most general condition. To derive the
correct Ward Identities we only need that just the {\em
combination} of action and measure is invariant. In this subsection
we want to discuss the more general case in which the set of
transformations (69) still generate a symmetry of the combination
of measure and action, but not of each individually. If we
insist on a solution of the form (68), then the other property
that is required, $\langle c^A\phi^*_B\rangle = -i\hbar\delta^A_B$,
follows automatically.

It thus remains to be found under what conditions the combination
of the action and the measure remain invariant under the BRST
Schwinger-Dyson symmetry. With an action $S_{ext}$ of the form
(68), we get
\begin{eqnarray}
\delta S_{ext} &=& \frac{\delta^r S^{(BV)}}{\delta\phi^A}c^A +
\frac{\delta^r S^{(BV)}}{\delta\phi^*_A}\left(-\frac{\delta^l
S_{ext}}{\delta\phi^A}\right) - \frac{\delta^r(\phi^*_Ac^A)}
{\delta\phi^*_B}\left(-\frac{\delta^l S_{ext}}{\delta\phi^A}
\right) \cr
&=& \frac{\delta^r S^{(BV)}}{\delta\phi^A}c^A +
\frac{\delta^r S^{(BV)}}{\delta\phi^*_A}\left(-\frac{\delta^l
S^{(BV)}}{\delta\phi^A}\right) - (-1)^{\epsilon_A+1}c^A
\left(-\frac{\delta^l S^{(BV)}}{\delta\phi^A}\right) \cr
&=& -\frac{\delta^r S^{(BV)}}{\delta\phi^*_A}\frac{\delta^l
S^{(BV)}}{\delta\phi^A} ~=~ \frac{1}{2}(S^{(BV)},S^{(BV)})~.
\end{eqnarray}

We will still assume that we are integrating over a flat euclidean
measure for the fundamental field $\phi^A$. This measure is formally
invariant under the transformation (69). However, for a corresponding
flat euclidean measure for $\phi^*_A$, the Jacobian of the
transformation (69) will in general be different from unity. As we
already discussed above, the Jacobian equals
\beq
J = 1 - \frac{\delta^r}{\delta\phi^*_A}\left(\frac{\delta^l S_{ext}}
{\delta\phi^A(x)}\mu\right)~.
\eeq

Thus to demand that the combination of measure and action remains
invariant, we must in general require that
\beq
\frac{1}{2}(S_{ext},S_{ext}) =
-\frac{\delta^r S_{ext}}{\delta\phi^A}c^A
+ i\hbar \Delta S_{ext} ~,
\eeq
which, assuming the form (68) -- since we know that this is sufficient
to guarantee the correct Schwinger-Dyson equations -- reduces to
the quantum Master Equation of Batalin and Vilkovisky:
\beq
\frac{1}{2}(S^{(BV)},S^{(BV)}) =  i\hbar \Delta S^{(BV)}~.
\eeq

Let us emphasize that this equation follows even {\em before} possible
gauge fixings. It is required in order that the general
Schwinger-Dyson equations for the fundamental fields are satisfied,
and is not postulated on only the requirement that the final functional
integral be independent of the gauge-fixing function. However, gauge
independence of the functional integral upon the addition of a term
of the form $\delta\Psi[\phi]$ now follows straightforwardly, since
for a functional $\Psi$ that depends only on the fields $\phi$, we
have $\delta^2\Psi[\phi] = 0$.

\subsection{Quantum BRST}

In subsection A we noted that the usual BRST Schwinger-Dyson
symmetry acquires a ``quantum correction" if one insists on using
the formalism where the new ghost fields $c^A$ have been integrated
out of the path integral. As we saw already in the case of no gauge
symmetries, this deforms the BRST operator:
\beq
\delta \to \sigma = \delta - i\hbar\Delta ~.
\eeq
The notation is not entirely precise, because the operator $\delta$
on the right hand side of this equation of course equals the operator
$\delta$ on the left hand side only modulo those changes incurred
by integrating out the ghosts $c^A$. But we keep it like this
to avoid complications in the notation. After having
integrated out the ghosts $c^A$, the BRST operator $\delta$ will
become identical to the variation within the antibracket.

Since this quantum deformation involves the same operator $\hbar\Delta$
that in certain specific cases may modify the classical Master Equation,
one might be led to believe that these two issues are related, $i.e.$,
that the ``quantum BRST" operator should only be applied when there are,
(or as a consequence of having) quantum corrections in the full
gauge-fixed action. This is actually not the case, and we
therefore find it useful
to return briefly to the meaning of the quantum BRST operator, here
denoted by $\sigma$.

Let us again choose the simplest solution to the Master Equation of
the form (68). We emphasize that it is immaterial whether this
extended action $S_{ext}$ satisfies the classical or quantum Master
Equations. Since we are interested in seeing the effect of integrating
out the ghosts $c^A$, consider, as in subsection A, the expectation
value of the BRST variation of an arbitrary functional
$G = G[\phi^A,\phi^*_A]$:
\begin{eqnarray}
\langle \delta G[\phi,\phi^*] \rangle &=& {\cal{Z}}^{-1}\int
[d\phi][d\phi^*][dc] \delta G[\phi,\phi^*]\exp\left[\frac{i}{\hbar}
\left(S^{(BV)} - \phi^*_Ac^A\right)\right] \cr
&=& {\cal{Z}}^{-1}\int
[d\phi][d\phi^*][dc]\left\{\frac{\delta^rG}{\delta\phi^A}c^A
+ \frac{\delta^rG}{\delta\phi^*_A}\left(-\frac{\delta^l S_{ext}}
{\delta\phi^A}\right)\right\}
\exp\left[\frac{i}{\hbar}\left(S^{(BV)} - \phi^*_Ac^A\right)\right]
\cr
&=& {\cal{Z}}^{-1}\int
[d\phi][d\phi^*]\left\{\frac{\delta^rG}{\delta\phi^A}(i\hbar)
\frac{\delta^l}{\delta\phi^*_A}\delta(\phi^*)
- \frac{\delta^rG}{\delta\phi^*_A}\frac{\delta^l S^{(BV)}}
{\delta\phi^A}\delta(\phi^*)\right\}
\exp\left[\frac{i}{\hbar}S^{(BV)}\right] \cr
&=& {\cal{Z}}^{-1}\int
[d\phi][d\phi^*]\delta(\phi^*)\left\{\frac{\delta^rG}{\delta\phi^A}
\frac{\delta^l S^{(BV)}}{\delta\phi*_A} + (i\hbar)(-1)^{\epsilon_A}
\frac{\delta^r\delta^rG}{\delta\phi*_A\delta\phi^A}
- \frac{\delta^rG}{\delta\phi^*_A}\frac{\delta^l S^{(BV)}}
{\delta\phi^A}\right\} \cr
&& \times \exp\left[\frac{i}{\hbar}S^{(BV)}\right] \cr
&=& \langle (G,S^{(BV)}) - i\hbar\Delta G \rangle~.
\end{eqnarray}
The derivation given here corresponds to the path integral before
gauge fixing, but it goes through in entirely the same manner in
the gauge-fixed case. (The only difference is that the relevant
$\delta$-function reads $\delta(\phi^* - \delta^r\Psi/\delta
\phi)$ instead of $\delta(\phi^*)$; this does not affect the
manipulations above).

The emergence of the ``quantum correction" in the BRST operator
is thus completely independent of the particular solution
$S^{(BV)}[\phi,\phi^*]$; it must always be included when one uses
the formalism in which the ghosts $c^A$ have been integrated out.
The quantum BRST operator $\sigma$ is unusual, because it
appears only after functional manipulations inside the path
integral.

Since by construction the partition function is invariant under
$\delta$ (when keeping the ghosts $c^A$) and $\sigma$ (after
having integrated out these ghosts), it follows that all expectation
values involving these operators vanish:
\beq
\langle \delta G[\phi,\phi^*] \rangle = 0
\eeq
when keeping $c^A$, and
\beq
\langle \sigma G[\phi,\phi^*] \rangle = 0
\eeq
when the $c^A$ have been integrated out.

This of course holds for the action as well:
\beq
\langle \delta S_{ext} \rangle = 0~;~~~~\langle \sigma S^{(BV)}
\rangle =  0~.
\eeq

The first of these equations is trivially satisfied when $S_{ext}$
satisfies the classical Master Equation, because then the variation
$\delta S_{ext}$ itself vanishes. This equation is then only
non-trivially satisfied when $\Delta S_{ext} \neq 0$.

Since the two operations $\delta$ and $\sigma$ are equivalent in
the precise sense given above, the
same considerations should apply to the second equation.
Indeed it does: When $S^{(BV)}$ satisfies the classical Master
Equation, $\sigma S^{(BV)} = 0$ at the operator level, while
that equation is satisfied only in terms of expectation values when
$\Delta S^{(BV)} \neq 0$.

Note that when $\Delta S_{ext} \neq 0$ (or $\Delta S^{(BV)} \neq 0$),
the quantum action is neither invariant under $\delta$ nor $\sigma$.
The action precisely has to remain non-invariant in order to cancel
the non-trivial contribution from the measure in that case. This is
the origin of the factor 1/2 difference between the quantum Master
Equation
\beq
\frac{1}{2}(S^{(BV)},S^{(BV)}) - i\hbar\Delta S^{(BV)} = 0
\eeq
and the operator $\sigma$ (when acting on $S^{(BV)}$):
\beq
\langle (S^{(BV)},S^{(BV)}) - i\hbar\Delta S^{(BV)} \rangle = 0 ~.
\eeq
The combination of these two equations yields the new identities
\beq
\langle \Delta S^{(BV)} \rangle = 0~,~~~ \langle (S^{(BV)},
S^{(BV)}) \rangle = 0 ~
\eeq
which can also be verified directly using the path integral.

The operator $\delta$ defines a BRST cohomology only on the
subspace of fields $\phi^A$; it is only nilpotent on that
subspace. The operator $\sigma$ is nilpotent in general:
$\sigma^2 = 0$ (a consequence of having performed partial
integrations in deriving it). However, the two operators
share the same physical content.

\section{Generalizations of the Batalin-Vilkovisky formalism}
\noindent
In order to see how the conventional antibracket formalism of Batalin and
Vilkovisky can be generalized, it is important to have a
fundamental principle from which this formalism can be derived. As has
been discussed in the previous sections, this
principle is that Schwinger-Dyson BRST symmetry must be
imposed on the full path integral.

\noi
Schwinger-Dyson BRST symmetry can be derived from the local symmetries
of the given path integral measure. When the measure is flat, the relevant
symmetry is that of local shifts, and the resulting Schwinger-Dyson BRST
symmetry leads directly to a quantum Master Equation on the action $S$
which is exponentiated inside the path integral. This action depends
on two new sets of ghosts and antighosts, $c^A$ and $\phi^*_A$ \cite{aldam1}.
The conventional Batalin-Vilkovisky formalism for an action $S^{BV}$
follows if one substitutes $S[\phi,\phi^*,c] = S^{BV}[\phi,\phi^*]
-\phi^*_Ac^A$ and integrates out the ghosts $c^A$.
The so-called ``antifields'' of the Batalin-Vilkovisky formalism are
simply the Schwinger-Dyson BRST antighosts $\phi^*_A$ \cite{aldam1}.

\noi
It is of interest to see what happens if one abandons\footnote{See the
2nd reference in \cite{aldam2}. This is related to the covariant formulations
of the antibracket formalism \cite{covariant}.} the assumption
of flat measures for the fields $\phi^A$, and if one does not restrict
oneself to local transformations that leave the functional measure
invariant. Some steps in this direction were recently taken in ref.
\cite{aldam2}. One here exploits the reparametrization invariance encoded
in the path integral by performing field transformations $\phi^A =
g^A(\phi',a)$ depending on new fields
$a^i$. It is natural to assume that these transformations form a
group, or more precisely, a quasigroup \cite{Bat}. The objects
\beq
u^A_i ~\equiv~ \left.\frac{\delta^r g^A}{\delta a^i}\right|_{a=0}
\eeq
are gauge generators of this group. They satisfy
\beq
\frac{\delta^r u^A_i}{\delta\phi^B}u^B_j - (-1)^{\epsilon_i\epsilon_j}
\frac{\delta^r u^A_j}{\delta\phi^B}u^B_i = - u^A_k U^k_{ij} ~,\label{u}
\eeq
where the $U^k_{ij}$ are structure ``coefficients'' of the group. They
are supernumbers with the property
\beq
U^k_{ij} = - (-1)^{\epsilon_i\epsilon_j}U^k_{ji} ~.
\eeq

\noi
In ref. \cite{aldam2}, specializing to compact supergroups for which
$(-1)^{\epsilon_i}U^i_{ij} = 0$, the following $\Delta$-operator was derived:
\beq
\Delta G \equiv (-1)^{\epsilon_i}\left[\frac{\delta^r}{\delta\phi^A}
\frac{\delta^r}{\delta\phi^*_i}G\right]u^A_i + \frac{1}{2}(-1)^{\epsilon_i+1}
\left[\frac{\delta^r}{\delta\phi^*_j}\frac{\delta^r}{\delta\phi^*_i}G\right]
\phi^*_k U^k_{ji} ~. \label{rank-one-delta}
\eeq
When the coefficients $U^k_{ij}$ are constant, this $\Delta$-operator
is nilpotent: $\Delta^2 = 0$. As noted by Koszul \cite{Koszul}, and
rediscovered by Witten \cite{Witten},
one can define an antibracket $(F,G)$ by the rule
\beq
\Delta(FG) = F(\Delta G) + (-1)^{\epsilon_G}(\Delta F)G
+ (-1)^{\epsilon_G}(F,G) ~. \label{abdef}
\eeq
Explicitly, for the case above, this leads to the following new
antibracket \cite{aldam2}:
\beq
(F,G) \equiv (-1)^{\epsilon_i(\epsilon_A+1)}\frac{\delta^r F}
{\delta\phi^*_i}u^A_i\frac{\delta^l G}{\delta\phi^A} - \frac{\delta^r F}
{\delta\phi^A}u^A_i\frac{\delta^l G}{\delta\phi^*_i} + \frac{\delta^r F}
{\delta\phi^*_i}\phi^*_k U^k_{ij}\frac{\delta^l G}{\delta\phi^*_j}
\eeq

\noi
This antibracket is statistics-changing,
$\epsilon((F,G)) = \epsilon(F) + \epsilon(G) + 1$, and has the following
properties:
\begin{eqnarray}
(F,G) &=& (-1)^{\epsilon_F\epsilon_G+\epsilon_F+\epsilon_G}(G,F)
\label{exchange} \\
(F,GH) &=& (F,G)H + (-1)^{\epsilon_G(\epsilon_F+1)}G(F,H) \cr
(FG,H) &=& F(G,H) + (-1)^{\epsilon_G(\epsilon_H+1)}(F,H)G \label{Leibniz}
\\
0 &=& (-1)^{(\epsilon_F+1)(\epsilon_H+1)}(F,(G,H))
+ {\mbox{\rm cyclic perm.}} ~.\label{Jacobi}
\end{eqnarray}
Furthermore,
\beq
\Delta (F,G) = (F,\Delta G) - (-1)^{\epsilon_G}(\Delta F,G) ~.\label{dfg}
\eeq

\noi
The $\Delta$ given in eq. (\ref{rank-one-delta}) is clearly
a non-Abelian generalization of the conventional $\Delta$-operator of the
Batalin-Vilkovisky formalism.

\noi
We shall now show how to extend this construction to the general
case of non-linear and open algebras. Recently, interest in
more complicated algebras such as strongly homotopy Lie algebras
\cite{Stasheff} has arisen in the context of string field theory
\cite{Zwiebach}.

\noi
Consider the quantized Hamiltonian BRST operator $\Omega$ for first-class
constraints with an arbitrary, possibly open,
gauge algebra \cite{BF}.\footnote{For
a comprehensive review of the classical Hamiltonian BRST formalism, see,
$e.g.$, ref. \cite{Henneaux}.} Apart from a set of phase space operators
$Q^i$ and $P_i$, introduce a ghost pair $\eta^i,
{\cal P}_i$. They have Grassmann parities $\epsilon(\eta^i) =
\epsilon({\cal P}_i) = \epsilon(Q^i) + 1 \equiv \epsilon_i + 1$, and are
canonically conjugate with respect to the usual graded commutator:
\beq
[\eta^i,{\cal P}_j] =  \eta^i{\cal P}_j - (-1)^{(\epsilon_i+1)(\epsilon_j+1)}
{\cal P}_j\eta^i = i\delta^i_j ~.\label{supercom}
\eeq
In addition $[\eta^i,\eta^j] = [{\cal P}_i,{\cal P}_j] = 0$.
The quantum mechanical BRST operator can then be written in the form of
a ${\cal P}\eta$ normal-ordered expansion in powers of the ${\cal P}$'s
\cite{BF}:
\beq
\Omega = G_i\eta^i + \sum_{n=1}^{\infty}{\cal P}_{i_{n}}\cdots
{\cal P}_{i_{1}} U^{i_{1}\cdots i_{n}} ~.\label{Omega}
\eeq
Here
\beq
U^{i_{1}\cdots i_{n}} = \frac{(-1)^{\epsilon^{i_{1}\cdots i_{n-1}}_{j_{1}
\cdots j_{n}}}}{(n+1)!}U^{i_{1}\cdots i_{n}}_{j_{1}
\cdots j_{n+1}}\eta^{j_{n+1}}\cdots\eta^{j_{1}} ~,
\eeq
and the sign factor is defined by:
\beq
\epsilon^{i_{1}\cdots i_{n-1}}_{j_{1}
\cdots j_{n}} = \sum_{k=1}^{n-1}\sum_{l=1}^{k}\epsilon_{i_{l}} +
\sum_{k=1}^n \sum_{l=1}^k \epsilon_{j_{l}} ~.
\eeq
The $U^{i_{1}\cdots i_{1}}_{j_{n}\cdots j_{n+1}}$'s are generalized
structure coefficients. For rank-1 theories the expansion ends with
the 2nd term, involving the usual Lie algebra structure coefficients
$U^k_{ij}$. The number of terms that must be included in the expansion
of eq. (\ref{Omega}) increases with the rank. By construction
$\Omega^2 = 0$.

\noi
The functions $G_i$ appearing in eq. (\ref{Omega}) are the constraints.
In the quantum case they satisfy the constraint algebra
\beq
[G_i,G_j] ~=~ iG_k U^k_{ij} ~.
\eeq
We choose these to be the ones associated with motion on the
supergroup manifold defined by the transformation $\phi^A =
g^A(\phi',a)$.

\noi
When considering representations of the (super) Heisenberg algebra
(\ref{supercom}), one normally chooses the operators to act to the
right. Thus, in the ghost coordinate representation we could take
\beq
{\cal P}_j = i (-1)^{\epsilon_j}\frac{\delta^l}{\delta\eta^j} ~,
\eeq
and similarly in the ghost momentum representation we could take
\beq
\eta^j = i \frac{\delta^l}{\delta {\cal P}_j} ~.
\eeq

\noi
On the other hand, the most convenient representation of the constraint
$G_j$ is \cite{Bat}
\beq
\lG_j = - i \frac{\ldr}{\delta \phi^A}u^A_j ~,
\eeq
which involves a right-derivative {\em acting to the left}.
Using eq. (\ref{u}), $\lG_j$ is seen to satisfy
\beq
[\lG_i,\lG_j] = i \lG_k U^k_{ij} ~.
\eeq

\noi
Since we wish $\Omega$ of eq. (\ref{Omega}) to act in a definite way,
we choose representations of the (super) Heisenberg algebra (\ref{supercom})
that involve operators acting to the left as well. These are
\beq
\lP_j = i \frac{\ldr}{\delta\eta^j}
\eeq
in the ghost coordinate representation, and
\beq
\leta_j = i (-1)^{\epsilon_j}\frac{\ldr}{\delta{\cal P}_j}
\eeq
in the ghost momentum representation.
Inserting these operators into eq. (\ref{Omega}) will give the corresponding
BRST operator $\lOmega$ acting to the left. We now identify the ghost
momentum ${\cal P}_j$ with the Lagrangian
antighost (``antifield'') $\phi^*_j$.

\noi
As a special case, consider
the operator $\lOmega$ in the case of an ordinary rank-1 super Lie algebra
for which $(-1)^{\epsilon_i}U^i_{ij} = 0$.
In the ghost momentum representation $\lOmega$ takes the form
\beq
\lOmega = (-1)^{\epsilon_i}\frac{\ldr}{\delta\phi^A}u^A_i\frac{\ldr}
{\delta\phi^*_i}
- \frac{1}{2}(-1)^{\epsilon_j}\phi^*_k U^k_{ij}\frac{\ldr}{\delta
\phi^*_j}\frac{\ldr}{\delta\phi^*_i} ~.
\eeq
One notices that the $\lOmega$ of the above equation
coincides with our non-Abelian $\Delta$-operator of eq. (
\ref{rank-one-delta}). In detail:
\beq
\Delta F ~\equiv~ F\lOmega ~. \label{delta-omega}
\eeq

\noi
For a rank-0 algebra -- the Abelian case --  we get,
with the same identification,
\beq
\lOmega = (-1)^{\epsilon_A}\frac{\ldr}{\delta\phi^A}\frac{\ldr}{
\delta\phi^*_A} ~.
\eeq
The associated $\Delta$-operator, defined through eq. (\ref{delta-omega})
is seen to agree with the $\Delta$ of the conventional
Batalin-Vilkovisky formalism eq.(9).

\noi
We define the general $\Delta$-operator through the
identification (\ref{delta-omega}) and the complete expansion
\beq
\lOmega = (-1)^i\frac{\ldr}{\delta\phi^A}u^A_i\frac{\ldr}{\delta\phi^*_i}
+ \sum_{n=1}^{\infty}\phi^*_{i_{n}}\cdots
\phi^*_{i_{1}} \lU^{i_{1}\cdots i_{n}} ~.\label{leftOmega}
\eeq
Here
\beq
\lU^{i_{1}\cdots i_{n}} = \frac{(-1)^{\epsilon^{i_{1}\cdots i_{n-1}}_{j_{1}
\cdots j_{n}}}}{(n+1)!}(i)^{n+1}(-1)^{\epsilon_{j_{1}} + \cdots
+ \epsilon_{j_{n+1}}}U^{i_{1}\cdots i_{n}}_{j_{1}
\cdots j_{n+1}}\frac{\ldr}{\delta\phi^*_{j_{n+1}}}\cdots\frac{\ldr}{
\delta\phi^*_{j_{1}}} ~.
\eeq
By construction we then have $\Delta^2 = 0$.

\noi
It is quite remarkable that the above derivation, based on Hamiltonian
BRST theory in the operator language,
has a direct counterpart in the Lagrangian path integral.
The two simplest cases, that of rank-0 and rank-1 algebras have been
derived in detail in the Lagrangian formalism in ref. \cite{aldam2}.
It is intriguing that completely different manipulations
(integrating out the corresponding ghosts $c^i$, and
partial integrations inside the functional integral) in the Lagrangian
framework leads to these quantized Hamiltonian BRST operators. The rank-0
case, that of the conventional Batalin-Vilkovisky formalism, corresponds
to the gauge generators
\beq
\lG_A = -i \frac{\ldr}{\delta\phi^A} ~.
\eeq
These are generators
of translations: when the functional measure is flat,
the Schwinger-Dyson BRST symmetry is generated
by local translations. The non-Abelian generalizations correspond to
imposing different symmetries as BRST symmetries in the path integral
\cite{aldam2}.

\noi
These non-Abelian BRST operators $\lOmega$ can be Abelianized by canonical
transformations involving the ghosts \cite{BF1}, but the significance of
this in the present context is not clear. Since in general the number
of ``antifields'' $\phi^*_i$ will differ from that of the fields
$\phi^A$, it is obvious that $u^A_i$ in general will be non-invertible.
Even when the number of antifields matches that of fields, the
associated matrix $u^A_B$ may be non-invertible (``degenerate'').\footnote{
In the special case where $u^A_B$ is invertible, the transformation
$\phi^*_A \to \phi^*_B(u^{-1})^B_A$ makes the corresponding $\Delta$-operator
Abelian \cite{us1}, but we are not interested in that case here. See
also refs. \cite{BT,Nersessian}.}

\noi
Having the general $\Delta$-operator available,
the next step consists in extracting
the associated antibracket. By the definition (\ref{abdef}), this
antibracket measures the failure of $\Delta$ to be a derivation. When
$\Delta$ is a second-order operator, the antibracket so defined will
itself obey the derivation rule (\ref{Leibniz}). For higher-order
$\Delta$-operators this is no longer the case. The antibracket will
then in all generality only obey the much weaker relation
\beq
(F,GH) = (F,G)H - (-1)^{\epsilon_G}F(G,H) + (-1)^{\epsilon_G}(FG,H) ~.
\eeq
The relation (\ref{dfg}) also holds in all generality.
When the $\Delta$-operator is of order three or higher, the antibracket
defined by (\ref{abdef}) will not only fail to be a derivation, but
will also violate the Jacobi identity (\ref{Jacobi}).

\noi
For higher-order $\Delta$-operators one can, as explained by Koszul
\cite{Koszul}, use the failure of the antibracket to be a derivation to
define {\em higher antibrackets}. These are Grassmann-odd analogues of
Nambu brackets \cite{Nambu,Takhtajan}. The construction is most
conveniently done in an iterative procedure, starting with the
$\Delta$-operator itself \cite{Koszul,Akman}. To this end, introduce
objects $\Phi^n_{\Delta}$ which are defined as follows:\footnote{Note
that our definitions differ slightly from ref. \cite{Koszul,Akman}
due to our $\Delta$-operators being based on right-derivatives, while those
of ref. \cite{Koszul,Akman} are based on left-derivatives.}
\begin{eqnarray}
\Phi^1_{\Delta}(A) &=& (-1)^{\epsilon_A}\Delta A \cr
\Phi^2_{\Delta}(A,B) &=& \Phi^1_{\Delta}(AB) - \Phi^1_{\Delta}(A) B
- (-1)^{\epsilon_A} A\Phi^1_{\Delta}(B) \cr
\Phi^3_{\Delta}(A,B,C) &=& \Phi^2_{\Delta}(A,BC) - \Phi^2_{\Delta}(A,B)C
- (-1)^{\epsilon_B(\epsilon_A+1)}B\Phi^2_{\Delta}(A,C) \cr
\cdot && \cdot \cr \cdot && \cdot \cr \cdot && \cdot \cr
\Phi^{n+1}_{\Delta}(A_1,\ldots,A_{n+1}) &=& \Phi^n_{\Delta}(A_1,
\ldots,A_nA_{n+1}) - \Phi^n_{\Delta}(A_1,\ldots,A_n)A_{n+1}\cr &&
- (-1)^{\epsilon_{A_{n}}(\epsilon_{A_{1}} + \cdots +\epsilon_{A_{n-1}} +1)}
A_n\Phi^n_{\Delta}(A_1,\ldots,A_{n-1},A_{n+1}) ~.
\end{eqnarray}

\noi
The $\Phi^n_{\Delta}$'s define the higher antibrackets. For example,
the usual antibracket is given by
\beq
(A,B) \equiv (-1)^{\epsilon_A}\Phi^2_{\Delta}(A,B) ~.
\eeq
The iterative procedure clearly stops at the first bracket that acts like
a derivation. For example, the ``three-antibracket'' defined by
$\Phi^3_{\Delta}(A,B,C)$ directly measures the failure of $\Phi^2_{\Delta}$
to act like a derivation. But more importantly, it also
measures the failure of the usual antibracket to satisfy
the graded Jacobi identity:
\begin{eqnarray}
\sum_{\mbox{\rm cycl.}}(-1)^{(\epsilon_A+1)(\epsilon_C+1)}(A,(B,C))
&=& (-1)^{\epsilon_A(\epsilon_C+1)+\epsilon_B+\epsilon_C}\Phi^1_{\Delta}(
\Phi^3_{\Delta}(A,B,C)) \cr
&&+ \sum_{\mbox{\rm cycl.}}(-1)^{\epsilon_A(\epsilon_C
+1)+\epsilon_B+\epsilon_C}\Phi^3_{\Delta}(\Phi^1_{\Delta}(A),B,C) ~,
\end{eqnarray}
and so on for the higher brackets.

\noi
When there is an infinite number of higher antibrackets, the associated
algebraic structure is analogous to
a strongly homotopy Lie algebra $L_{\infty}$.
The $L_1$ algebra is then given by the nilpotent $\Delta$-operator,
the $L_2$ algebra is given by $\Delta$ and the usual antibracket,
the $L_3$ algebra by these two and the additional ``three-antibracket'',
etc. The set of higher antibrackets
defined above seems natural in closed string field theory
\cite{Zwiebach}, the corresponding $\Delta$-operator being
given by the string field BRST operator $Q$.

\noi
Having constructed the $\Delta$-operator (and its associated hierarchy
of antibrackets), it is natural to consider a quantum Master Equation of
the form
\beq
\Delta \exp\left[\frac{i}{\hbar}S(\phi,\phi^*)\right] = 0 ~.
\eeq
Using the properties of the $\Phi^{n}$'s defined above, we can write
this Master Equation as a series in the higher antibrackets,
\beq
\sum_{k=1}^{\infty}\left(\frac{i}{\hbar}\right)^k
\frac{\Phi^{k}(S,S,\ldots,S)}{k!} ~=~ 0 ~,\label{me}
\eeq
where each of the higher antibrackets $\Phi^{k}(S,S,\ldots,S)$ has $k$
entries. The series terminates at a finite order if the associated
BRST operator terminates at a finite order. For example, in the Abelian
case of shift symmetry the general equation (\ref{me}) reduces to
$i\hbar\Delta S - \frac{1}{2}(S,S) = 0$, the Master Equation of the
conventional Batalin-Vilkovisky formalism.

\noi
A solution $S$ to the general Master Equation (\ref{me}) is invariant under
deformations
\beq
\delta S = \sum_{k=1}^{\infty}\left(\frac{i}{\hbar}\right)^{k-1}
\frac{\Phi^{k}(\epsilon,S,S,\ldots,S)}{(k-1)!} ~,\label{Ssym}
\eeq
where again each $\Phi^{k}$ has $k$ entries, and $\epsilon$ is Grassmann-odd.
One can view this as the possibility of adding a BRST variation
\beq
\sigma\epsilon = \sum_{k=1}^{\infty}\left(\frac{i}{\hbar}\right)^{k-1}
\frac{\Phi^{k}(\epsilon,S,S,\ldots,S)}{(k-1)!} \label{mesym}
\eeq
to the action. Here $\sigma$ is the appropriately generalized ``quantum
BRST operator''.\footnote{For finite order, a rearrangement in terms of
increasing rather than decreasing orders of $\hbar$ may be more convenient.}
In the case of the Abelian shift symmetry, the above $\sigma$-operator
becomes $\sigma \epsilon = \Delta\epsilon + (i/\hbar)(\epsilon,S)$, which
precisely equals ($(i\hbar)^{-1}$ times) the quantum BRST operator of the
conventional Batalin-Vilkovisky formalism.

\noi
We note that the general Master Equation (\ref{me}) and the BRST symmetry
(\ref{Ssym}) has the same relation to closed string field theory
\cite{Zwiebach,Kugo} that the conventional Batalin-Vilkovisky Master
Equation and BRST symmetry has to open string field theory \cite{Witten}.
The r\^{o}le of the action $S$ is then played by the string field $\Psi$,
and the Master Equation (\ref{me}) is the analogue of the closed string
field equations. The symmetry (\ref{Ssym}) is then the analogue
of the closed string field theory gauge transformations.

\noi
The present definition of higher antibrackets suggests the existence
of an analogous hierarchy of Grassmann-even brackets based on a
supermanifold and a non-Abelian open algebra --  a natural generalization
of Poisson-Lie brackets. It is also interesting to
investigate the Poisson brackets and Nambu brackets
generated by the generalized antibrackets
and suitable vector fields $V$ anticommuting with the generalized
$\Delta$-operator (and in particular certain
Hamiltonian vector fields within the antibrackets),
as described in the case of the usual antibracket in ref. \cite{Nersessian1}.
We explore this in the next section.

\section{Poisson bracket and antibracket}

In this section we want to make an explicit connection between Poisson bracket
and the antibracket.

For this purpose, consider the canonical algebra:
\beqn
[x^i,x^j]=0 \\ 
{[p_i,p_j]}=0 \\
{[x^i,p^j]}=i\delta_{ij} 1
\eeqn

Now consider the non-abelian antibracket corresponding to a Lie algebra (see second term of 
equation (85)):
\beqn
(A,B)=\frac{\partial_r A}{\partial z_i^*}z_k^*U^k_{ij}\frac{\partial_l B}{\partial z_j^*}
\eeqn
We choose
$U^k_{ij}$  corresponding to the structure constants of the canonical algebra.
For each generator of the algebra we include an antifield, $z_i^*$ ($z_0^*$ is the antifield
associated to the generator $1$).

Now we introduce the operator $d$ ("exterior derivative"). For functions of
$z^i$ alone (i.e they do not depend on the antifields), it is:

\beqn
dA=A,_iz_i^*\\
dB=B,_iz_i^*
\eeqn

For the canonical algebra, we get:
\beqn
(dA,dB)=\{A,B\}z_0^*\\
\{A,B\}=\frac{\partial A}{\partial x^i}\frac{\partial B}{\partial p_i}-
\frac{\partial A}{\partial p_i}\frac{\partial B}{\partial x^i}
\eeqn
 We
see that $\{,\}$ is the usual Poisson bracket.

\subsection{Nambu bracket and generalizations}

In \cite{Nambu,Takhtajan} it is considered a possible generalization of the canonical formalism
of classical mechanics, where the evolution equation contains two (or more) Hamiltonians:
\beqn
\frac{d F}{dt}=[F,H_1,H_2]\\
{[G_1,G_2,G_3]}=\epsilon_{ijk}\frac{\partial G_1}{\partial x^i}\frac{\partial G_2}{\partial x^j}
\frac{\partial G_3}{\partial x^k}, \  i,j,k=1,2,3
\eeqn
$[G_1,G_2,G_3]$ is called a Nambu bracket.

Now we show how to obtain it (and its generalizations) from the higher antibrackets of \cite{aldam3}

Introduce:
\beqn
\Delta=\lambda\epsilon_{ijk}\frac{\partial }{\partial x_i^*}\frac{\partial }{\partial x_j^*}
\frac{\partial }{\partial x_k^*}, \  i,j,k=1,2,3\\
\Delta^2=0
\eeqn
For the 3-bracket we obtain:
\beqn
\Phi_3(A,B,C)=6\lambda(-1)^{\epsilon(B)}\epsilon_{ijk}\frac{\partial A }{\partial x_i^*}
\frac{\partial B }{\partial x_j^*}
\frac{\partial C}{\partial x_k^*}
\eeqn

Now, as in the previous subsection, choose:
\beqn
dG_1=G_{1,i}x_i^*\\
dG_2=G_{2,i}x_i^*\\
dG_3=G_{3,i}x_i^* \label{1}
\eeqn
the $G_k$ are functions of $x^j$ alone (they do not depend on the antifields).

Choosing $\lambda$ appropiately, we get:
\beqn
\Phi_3(dG_{1},dG_{2},dG_{3})=[G_1,G_2,G_3]\label{2}
\eeqn

It is clear  that the identities satisfied by $\Phi_3(A,B,C)$ imply
identities for $[G_1,G_2,G_3]$\cite{Takhtajan}. They, in turn, are a direct consequence of
$\Delta$ being nilpotent and of third order in the derivatives.

The most general Nambu bracket and its properties are obtained in a similar 
way, starting from a suitable nilpotent $\Delta$-operator and the 
identification eq.(\ref{1}) and eq.(\ref{2}.

\section{Conclusions}

In these lectures we have studied the Batalin-Vilkovisky(BV) method of 
quantization from the perspective of the BRST-Schwinger-Dyson symmetry.
This has led us directly to a generalization of the algebraic structure
behind BV, 
first to non-abelian
Lie algebras and later on, to general open algebras. New $\Delta$ operators
emerge quite naturally. From them, using the Koszul procedure, we have derived
a tower of n-brackets, a master equation and a corresponding invariance of the
master equation.

As a simple application of the new formalism, we have exhibit a connection between the 
non-abelian antibracket and the Poisson bracket. In addition to this, we have
explained how to get the Nambu brackets from a suitable $\Delta$ operator
combined with the Koszul procedure.

\section*{Acknowledgements}

The author wants to express his gratitude to the organizers of the
"VII Mexican School of Particles and Fields" and
"I Latin American Symposium on High Energy Physics" for a 
very pleasant stay at M\'erida. He also wants to thank the hospitality
of L.F. Urrutia at Universidad Nacional Aut\'onoma de M\'exico.

His work has been partially supported by Fondecyt \# 1950809 and a collaboration
Conacyt(M\'exico)-Conicyt(Chile).

\vspace{2cm}

\appendix{{\Large {\bf Appendix}}}

\vspace{0.3cm}

In this appendix we give some additional conventions, and
list some useful identities.

The Leibniz rules
for derivations of the left and right kind read
\beq
\frac{\delta^l(F\cdot G)}{\delta A} = \frac{\delta^lF}{\delta A} G
+ (-1)^{\epsilon_F\cdot\epsilon_A} F \frac{\delta^lG}{\delta A}
\eeq
and
\beq
\frac{\delta^r(F\cdot G)}{\delta A} = F \frac{\delta^rG}{\delta A}
+ (-1)^{\epsilon_G\cdot\epsilon_A} \frac{\delta^rF}{\delta A}G ~,
\eeq
where $A$ denotes a field (or antifield) of arbitrary Grassmann
parity $\epsilon_A$. Similarly, $\epsilon_F$ and $\epsilon_G$ are
the Grassmann parities of the functionals $F$ and $G$.

Actual variations, let us denote them by $\bar{\delta}$ in contrast
to the BRST transformations $\delta$ of the paper, are defined as
follows:
\beq
F[A+\bar{\delta}A] - F[A] \equiv \bar{\delta}F \equiv
\bar{\delta}A\frac{\delta^lF}{\delta A} \equiv \frac{\delta^rF}
{\delta A}\bar{\delta}A ~.
\eeq

The commutation rule of two arbitrary fields is
\beq
A\cdot B = (-1)^{\epsilon_A\cdot\epsilon_B}B\cdot A ~,
\eeq
and for actual variations one has the simple rule that
\beq
\bar{\delta}(F\cdot G) = (\bar{\delta}F)G + F(\bar{\delta}G)
\eeq
independent of the Grassmann parities $\epsilon_F$ and $\epsilon_G$.
The rules (122) and (123) in conjunction lead to the useful identity
\beq
\frac{\delta^l F}{\delta A} = (-1)^{\epsilon_A(\epsilon_F+1)}
\frac{\delta^r F}{\delta A} ~.
\eeq

The BRST variations we have worked with in this paper correspond to
right derivation rules. This is of course not imposed upon us, but it
is convenient if we wish to compare our expressions with those of

Batalin and Vilkovisky. It follows from requiring the actual
variations to be related to the BRST transformations by multiplication
of an anticommuting parameter $\mu$ {\em from the right}. This then
provides us with very helpful operational rules for the BRST
transformations $\delta$. In particular,
\beq
\bar{\delta}F \equiv (\delta F)\mu = \frac{\delta^r F}{\delta A}
\bar{\delta} A ~.
\eeq
Now, since
\beq
\delta F \equiv \frac{\delta^r \bar{\delta} F}{\delta\mu} ~,
\eeq
it follows that
\beq
\delta F = \frac{\delta^r F}{\delta A} \delta A~.
\eeq
{}From this it also follows directly that the BRST transformations
act as right derivations:
\beq
\delta(F\cdot G) = F(\delta G) + (-1)^{\epsilon_G}(\delta F)G ~.
\eeq

These are the basic rules that are needed for the manipulations in
the main text.

\end{document}